\documentclass[aps,prb,twocolumn,groupedaddress,showpacs,apsrev]{revtex4-1}
\usepackage[T1]{fontenc}
\usepackage{lmodern}
\usepackage[latin1]{inputenc}
\usepackage[swedish,english]{babel}
\usepackage{amsmath}
\usepackage{subfig}
\usepackage{amsfonts}
\usepackage{graphicx} 
\usepackage{comment}
\usepackage{xfrac}
\usepackage{color}
\usepackage{multirow}

\graphicspath{{./graphics/}}

\begin{document}

\title{Electronic structure and magnetic properties of L1$_0$ binary alloys}
 
\author{Alexander Edstr\"om}
\author{Jonathan Chico}
\author{Adam Jakobsson}
\author{Anders Bergman}
\author{Jan Rusz}
\affiliation{Department of Physics and Astronomy, Uppsala University, Box 516, 75121 Uppsala, Sweden}

\date{\today}

\begin{abstract}
We present a systematic study of the magnetic properties of L1$_0$ binary alloys FeNi, CoNi, MnAl and MnGa via two different density functional theory approaches. Our calculations show large magnetocrystalline anisotropies in the order $1~\text{MJ/m}^3$ or higher for CoNi, MnAl and MnGa while FeNi shows a somewhat lower value in the range $0.48 - 0.77 ~\text{MJ/m}^3$. Saturation magnetization values of $1.3~\text{MA/m}$, $1.0~\text{MA/m}$, $0.8~\text{MA/m}$ and $0.9~\text{MA/m}$ are obtained for FeNi, CoNi, MnAl and MnGa respectively. Curie temperatures are evaluated via Monte Carlo simulations and show $T_\text{C}=916~\text{K}$ and $T_\text{C}=1130~\text{K}$ for FeNi and CoNi respectively. For Mn-based compounds Mn-rich off-stoichiometric compositions are found to be important for the stability of a ferro or ferrimagnetic ground state with $T_\text{C}$ greater than $600~\text{K}$. The effect of substitutional disorder is studied and found to decrease both magnetocrystalline anisotropies and Curie temperatures in FeNi and CoNi.
\end{abstract}

\maketitle

Materials exhibiting a large saturation magnetization ($M_\text{s}$), high Curie temperature ($T_\text{C}$), as well as large magnetic anisotropy energy (MAE), are of great technological importance in a wide range of permanent magnet applications, from electric motors and generators to magnetic storage devices. L1$_0$ ordering of binary compounds is known to be able to significantly increase MAE relative to the disordered state and for certain materials, such as FePt, an enormous MAE in the order of $5~\text{MJ/m}^3$ is observed \cite{Kota2012, 0953-8984-16-48-019, PhysRevB.63.144409, PhysRevB.66.024413}. Large values for $M_\text{s}$ and $T_\text{C}$ can be obtained with cheap and abundant materials such as bcc Fe, while achieving large MAE is a challenge. Typically, large values of the MAE are obtained for materials containing heavy elements, such as platinum or rare-earths, providing strong spin-orbit coupling. Such elements are often scarcely available and thus expensive. Finding new materials, with large MAE, made from cheap and readily available elements is therefore a task of great technological importance. Certain L1$_0$ ordered binary compounds, such as FeNi\cite{Miura2013,Kota2012,Kojima2011,Neel1964,Kojima2012,Kotsugi2013}, CoNi\cite{Fukami2013}, MnAl\cite{Koch1960,Park2010,Nie2013, Coey2014} and MnGa\cite{Sakuma1998}, have been reported to exhibit large MAE without containing platinum or rare-earths, making them potentially interesting candidates for permanent magnet materials.

In this work, a thorough investigation is done into the electronic structure and magnetic properties of L1$_0$ structured binary compounds FeNi, CoNi, MnAl and MnGa. To the best of our knowledge, first principles all-electron electronic structure calculations including full-potential effects have not been presented in the literature for all these compounds yet. Furthermore, all three of the important permanent magnet properties $M_\text{s}$, $T_\text{C}$ and MAE are adressed for all of the compounds. In addition to this, substitutional disorder and off-stoichiometric compositions are investigated. 

\begin{figure}[hbt!]
	\centering 
	\includegraphics[trim = 25mm 25mm 25mm 15mm, width=0.45\textwidth]{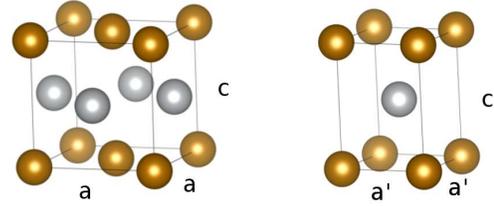}
	\caption{Two different unit cells of the L1$_0$ structure. $a'=\frac{a}{\sqrt{2}}$. }
	\label{fig:struct}
\end{figure}

\begin{figure}[hbt!]
	\centering 
	\includegraphics[width=0.37\textwidth]{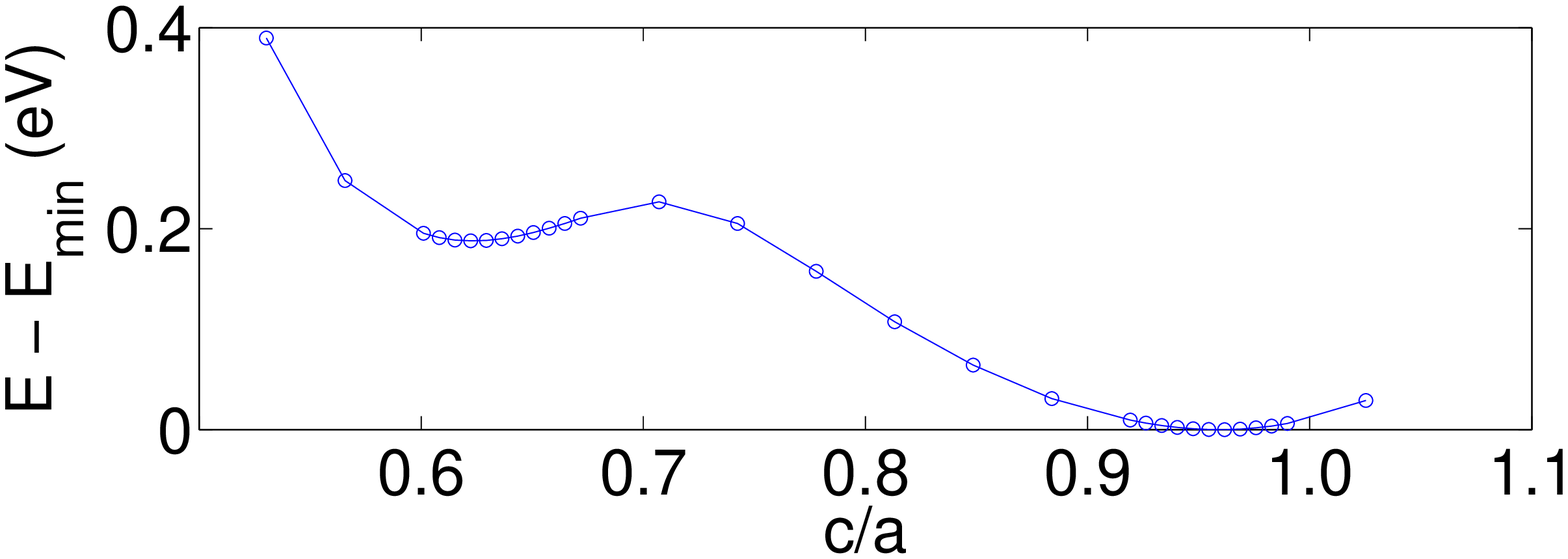}
	\caption{Difference in total energy and total energy of the equilibrium structure as function of $\frac{c}{a}$, varied under constant volume, for MnGa. }
	\label{fig:MnGaca}
\end{figure}

Three different computational methods were utilized in the calculations behind this work. First, two density functional theory (DFT) implementations, namely full-potential all-electron code WIEN2k\cite{Blaha2001} with linearized augmented plane wave basis functions and the Munich spin polarized relativistic Korringa-Kohn-Rostoker (SPR-KKR) package\cite{sprkkr, Ebert2011} were used, both with the generalized gradient approximation\cite{PhysRevLett.77.3865} for the exchange-correlation potential, to calculate ground state properties of the investigated systems. Later, Monte Carlo (MC) simulations of the Heisenberg hamiltonian were performed, using the Uppsala Atomistic Spin Dynamics (UppASD)\cite{Skubic2008} method, with exchange parameters calculated, via the method of Liechtenstein et al.\cite{Liechtenstein1984,Liechtenstein1986}, in SPR-KKR. Results of these calculations are shown in Table~\ref{table1}. The L1$_0$ structure can be described by either a bct or fct-like unit cell as illustrated in Fig.~\ref{fig:struct}. The smaller bct-like unit cell is used as input for calculations, as it allows lower computational cost due to a smaller basis, while Table~\ref{table1} contains lattice parameters describing the fct-like cell, as it is commonly used and gives a $c/a$-ratio better describing the deviation from a cubic structure. The lattice parameters were evaluated by total energy minimization in WIEN2k and used as input for all further calculations. In the case of MnGa a double minimum is observed in the total energy as function of $\frac{c}{a}$ as shown in Fig.~\ref{fig:MnGaca}. The data for MnGa shown in Table~\ref{table1} is for the more stable structure, with larger $\frac{c}{a}$, which shows a rather large uniaxial MAE, in contrast to the structure in the local minimum, which reveals a smaller in-plane anisotropy. The MAE was evaluated using the torque method\cite{Staunton2006, Ebert2011} in SPR-KKR and total energy difference calculations in WIEN2k. 160000 \textbf{k}-vectors and 40 energy points were used in SPR-KKR and basis functions up to $l=3$ were included. In WIEN2k, 20000 or more \textbf{k}-vectors were used, the smallest muffin-tin radius times maximum $\mathbf{k}$-vector was set to $R_{\text{MT}}K_{\text{max}}=9$ or higher and Brillouin-zone integration was performed using the modified tetrahedron method\cite{Blochl1994}. 

Magnetocrystalline anisotropy is a relativistic phenomenon due to the spin-orbit coupling (SOC). The two DFT methods used differ in the way they take relativistic effects, in general, and SOC, in particular, into account. WIEN2k does a fully relativistic treatment of the core electrons but a scalar relativistic approximation for the valence electrons with SOC included as a perturbation \cite{Schwarz2002}. This should be a very accurate method for 3d metals and has been shown to yield good results even for significantly heavier elements\cite{Koelling1977, Shick2005}. The SPR-KKR method, on the other hand, deals with relativistic effects in all electrons via a fully relativistic four component Dirac formalism\cite{Ebert2011}.

The data in Table \ref{table1} show a good agreement between SPR-KKR and WIEN2k, although there is some minor disagreement in the MAE where SPR-KKR consistently yields a larger value. There are a number of reasons which can contribute to the difference in the MAE found from the two methods. One of the main possible reasons is that we did not take full-potential effects into account in the SPR-KKR calculations. Other reasons include that, as mentioned, relativistic effects are treated differently and also different basis functions are used to describe the Kohn-Sham orbitals. Furthermore, MAEs are typically relatively small energies orders of magnitude smaller than, for example, cohesive energies and hence difficult to obtain numerically with high accuracy. In view of this, the agreement between the two methods can be considered very good. The MAE has previously been calculated to $0.5~\text{MJ/m}^3$, $1.0~\text{MJ/m}^3$, $1.5~\text{MJ/m}^3$ and $2.6~\text{MJ/m}^3$ for FeNi, CoNi, MnAl and MnGa, respectively\cite{Miura2013,wu1999,Park2010,Sakuma1998}, consistent with the results presented here. Table \ref{table1} also contains experimental values for MAE, where available, for comparison. For CoNi and MnAl we see that the theoretical MAEs, both from SPR-KKR and WIEN2k are higher than reported experimental values. This is expected as experimental samples typically do not have perfect ordering and experiments are done at finite temperatures, factors which are known to reduce MAE\cite{Kota2012,0953-8984-16-48-019}. However, in the case of FeNi theoretical and experimental values are of similar magnitude even though perfectly ordered samples have not been synthesized. This might indicate that the theoretical values presented here are too low, possibly because these calculations ignore orbital polarisation corrections which have been reported to significantly increase MAE in FeNi \cite{PhysRevB.63.144409, Miura2013}.

\begin{table*}
\begin{tabular}{ | l | c | c | c | c | c | c | }
	    \hline
	    Quantity 							& FeNi & CoNi & MnAl & MnGa & Mn$_{1.14}$Al$_{0.86}$ & Mn$_{1.2}$Ga$_{0.8}$ 		\\ \hline
	    a (\AA) 							& 3.56 & 3.49 & 3.89 &  3.83  	& 3.89 		&  3.83		\\ 
	    c (\AA) 							& 3.58 & 3.60 & 3.49 &  3.69  	& 3.49 		&  3.69		\\ \hline
	    $m_\text{X}^{\text{W2k}} ~ (\mu_\text{B})$ 			& 2.69 & 1.77 & 2.33 &  2.56 	& - 		& -		\\ 
	    $m_\text{X}^{\text{kkr}} ~ (\mu_\text{B})$ 			& 2.73 & 1.75 & 2.49 &  2.74 	& 2.54/-3.41 	& 2.69/-3.40	\\ 
	    $m_\text{Y}^{\text{W2k}} ~ (\mu_\text{B})$ 			& 0.67 & 0.71 & -0.04 & -0.08 	& - 		& - 		\\ 
	    $m_\text{Y}^{\text{kkr}} ~ (\mu_\text{B})$ 			& 0.62 & 0.68 & -0.09 & -0.12 	& -0.10 	& -0.12		\\ 
	    $m_\text{tot}^{\text{W2k}} ~ (\text{MA/m})$ 		& 1.33 & 1.01 & 0.82 & 0.86 	& - 		& - 		\\ 
	    $m_\text{tot}^{\text{kkr}} ~ (\text{MA/m})$			& 1.37 & 1.03 & 0.84 & 0.90 	& 0.69		& 0.66 	 	\\ \hline
	    $E_{\text{MAE}}^{\text{W2k}} ~ (\mu \text{eV}/\text{f.u.})$ & 68.7 & 135.1 & 275.1 & 378.2 	& - 		& - 		\\ 
	    $E_{\text{MAE}}^{\text{kkr}} ~ (\mu \text{eV}/\text{f.u.})$ & 110.3 & 184.7 & 320.8 & 385.7 & 360.2		& 428.8 	\\ 
	    $E_{\text{MAE}}^{\text{W2k}} ~ (\text{MJ/m}^3)$ 		& 0.48 & 0.99 & 1.67 & 2.24 	& - 		& -	 	\\ 
	    $E_{\text{MAE}}^{\text{kkr}} ~ (\text{MJ/m}^3)$ 		& 0.77 & 1.35 & 1.95 & 2.28	& 2.18 		& 2.54		\\ 
	    $E_{\text{MAE}}^{\text{exp}} ~ (\text{MJ/m}^3)$ 		& 0.58~\cite{Mizuguchi2011} & 0.54~\cite{Fukami2013} & 1.37~\cite{Nie2013} & - & - & -	\\ \hline
	    $T_{\text{C}}^{\text{MFT}} ~ (\text{K})$ 			& 1107 & 1383 & -  & 107 & -  & -	\\ 
	    $T_{\text{C}}^{\text{MC}} ~ (\text{K})$ 			& 916 & 1130 & - & 80 & 670 & 690				\\ \hline
\end{tabular} 
\caption{Lattice parameters calculated using WIEN2k, magnetic moments and magnetic anisotropies calculated using WIEN2k and SPR-KKR as well as Curie temperatures calculated using mean field theory and UppASD Monte Carlo for L1$_0$ binary alloys FeNi, CoNi, MnAl and MnGa.}
\label{table1}
\end{table*}

\begin{figure}[hbt!]
	\centering 
	\subfloat[FeNi]{\label{fig:FeNiJ}\includegraphics[width=0.25\textwidth]{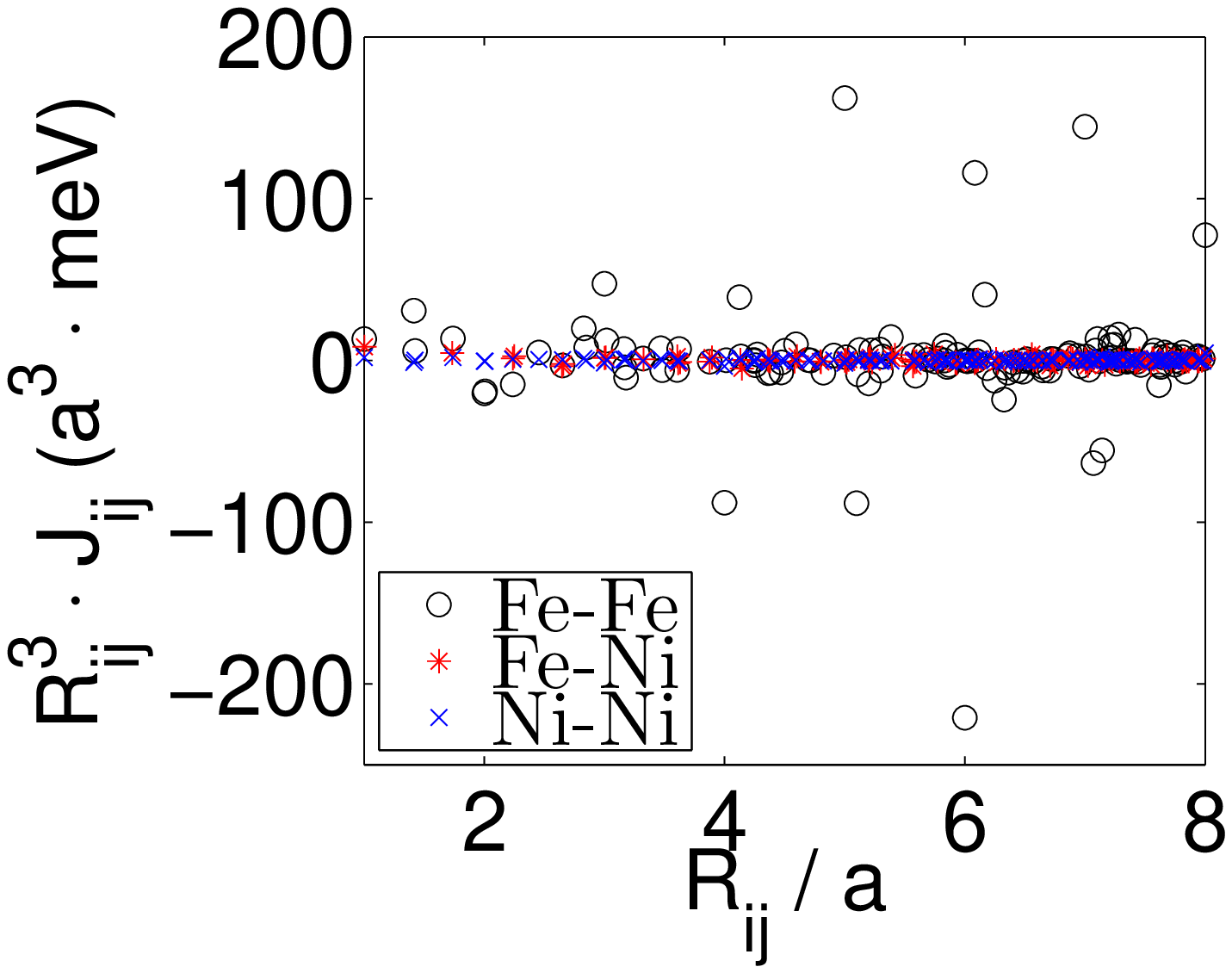}}
	\subfloat[CoNi]{\label{fig:CoNiJ}\includegraphics[width=0.25\textwidth]{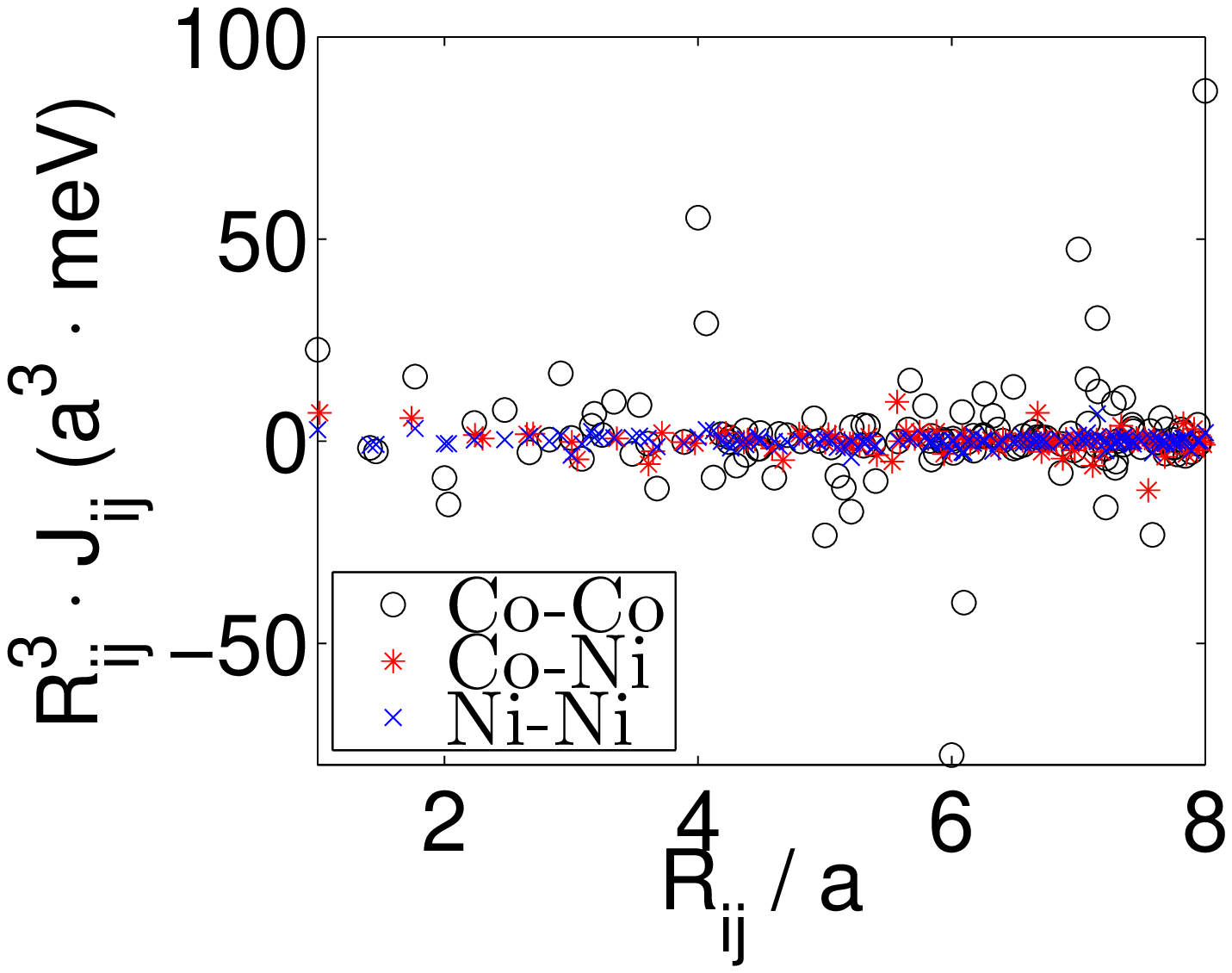}}
	\caption{Atomic distance dependence of exchange parameters $J_{ij}$.}
	\label{fig:Jxc}
\end{figure}

Exchange parameters, $J_{ij}$, were calculated in SPR-KKR and Fig.~\ref{fig:Jxc} shows how these vary with atomic distances for FeNi and CoNi. The $J_{ij}$ can be seen to decrease approximately as $R^{-3}$, as one would expect for metals with RKKY-type exchange interactions. These exchange parameters were used to calculate the Curie temperatures, presented in Table~\ref{table1}, via mean field theory (MFT) as well as MC simulations. MC Curie temperatures in the thermodynamic limit were evaluated by finite size scaling using the Binder cumulant method\cite{LandauBinder}. As expected, MFT overestimates $T_\text{C}$ compared to MC by around 20\%. Both Curie temperatures of $916~\text{K}$ and $1130~\text{K}$ for FeNi and CoNi are very high, which is suitable for permanent magnet applications. An MFT estimate of $T_\text{C}$ has previously been done to $1000\pm200~\text{K}$\cite{Lewis2014} for FeNi which is consistent with results presented here. The $J_{ij}$ are particularly large for Fe-Fe and Co-Co interactions, indicating that these elements contribute significantly to providing a high $T_\text{C}$ to the materials.

Real samples of L1$_0$ alloys do not exhibit perfect ordering and, for example, FeNi samples have been reported with long-range chemical order parameter around $S=0.48$~\cite{Kojima2012} ($S$ describes the fraction of atoms on the correct sublattice as $P = \frac{1}{2}(1+S)$). Disorder has been found to be important and have a negative effect on the MAE of FeNi as well as a number of other L1$_0$ materials\cite{Kota2012} and could also significantly affect $T_\text{C}$. Table~\ref{tablesubdis} shows the effect of some substitutional disorder on the MAE and $T_\text{C}$ of FeNi and CoNi. Calculations were perfomed on systems with one atomic position occupied by X$_{1-\eta}$Ni$_\eta$ and the other one by X$_{\eta}$Ni$_{1-\eta}$, with X=Fe or Co and $\eta$ up to $\eta=10\%$. Disorder was treated using the coherent potential approximation (CPA)\cite{Soven1967} in SPR-KKR. The data show how disorder causes a similar reduction of MAE, also in CoNi, as it does in FeNi and other L1$_0$ alloys. Also the $T_\text{C}$ of both FeNi and CoNi show a clear decrease with increasing disorder, although they still remain at high temperatures, well above room temperature. 

\begin{table}
\begin{tabular}{ |l|l|c|c|c| }
\hline
   & $\eta$ 	& 0\%	& 5\%				& 10\%			\\ \hline
\multirow{2}{*}{FeNi} 	 
 						& $E_{\text{MAE}}^{\text{kkr}} ~ (\mu \text{eV}/\text{f.u.})$ 	& 110.3 & 102.0 & 89.5 \\  
 						& $T_{\text{C}}^{\text{MC}} ~ (\text{K})$ 	& 916 	& 880 & 860 \\ \hline
\multirow{2}{*}{CoNi} 	& $E_{\text{MAE}}^{\text{kkr}} ~ (\mu \text{eV}/\text{f.u.})$ 	& 184.7 & 170.3 &	145.2		\\ 
 						& $T_{\text{C}}^{\text{MC}} ~ (\text{K})$  & 1130 & 940 & 935 \\ \hline
\end{tabular}
\caption{MAE and $T_\text{C}$ for FeNi and CoNi with substitutional disorder described by $\eta$.}
\label{tablesubdis}
\end{table}

It was recently suggested, based on experimental observations, that increasing the Fe-content in FeNi to Fe$_{1.2}$Ni$_{0.8}$ can increase MAE by around $30\%$\cite{Kojima2014}. SPR-KKR-CPA calculations failed to reproduce this result and rather indicated a reduction of MAE by around $10\%$ to $\text{MAE} = 98~\mu\text{eV/f.u.}$ in such a composition. Similarly, in Co$_{1.2}$Ni$_{0.8}$, the MAE was reduced to $141~\mu\text{eV/f.u.}$. Also the $T_\text{C}$ was reduced to $840~\text{K}$ and $1020~\text{K}$ in Fe$_{1.2}$Ni$_{0.8}$ and Co$_{1.2}$Ni$_{0.8}$ respectively. This can be understood from the exchange coupling parameters where there is a slight reduction in the strong positive parameters as one adds excess Fe or Co (not shown). 

\begin{figure}[hbt!]
	\centering 
	\subfloat[MnAl]{\label{fig:MnAlJ}\includegraphics[width=0.25\textwidth]{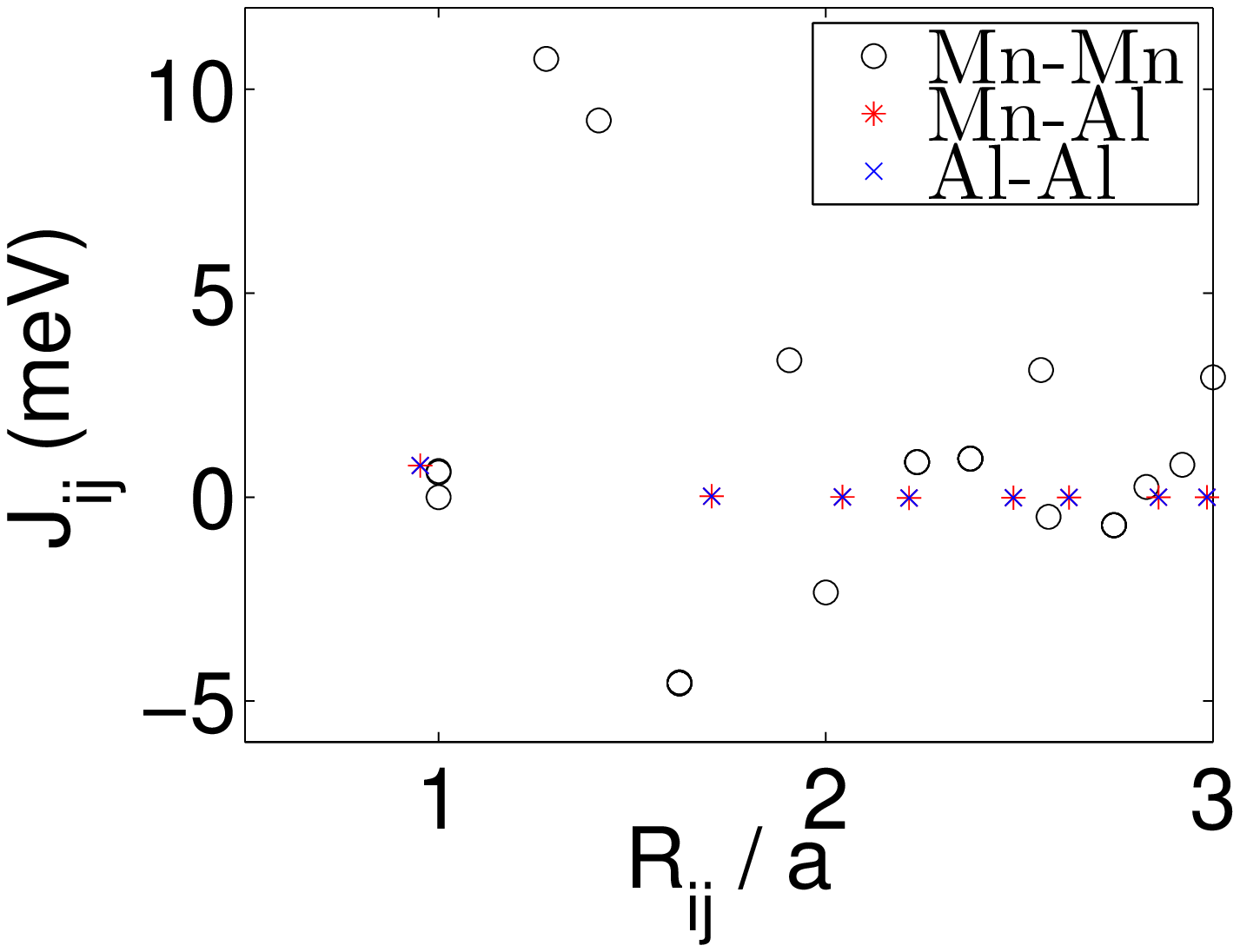}}
	\subfloat[Mn$_{1.14}$Al$_{0.86}$]{\label{fig:Mn114Al86J}\includegraphics[width=0.25\textwidth]{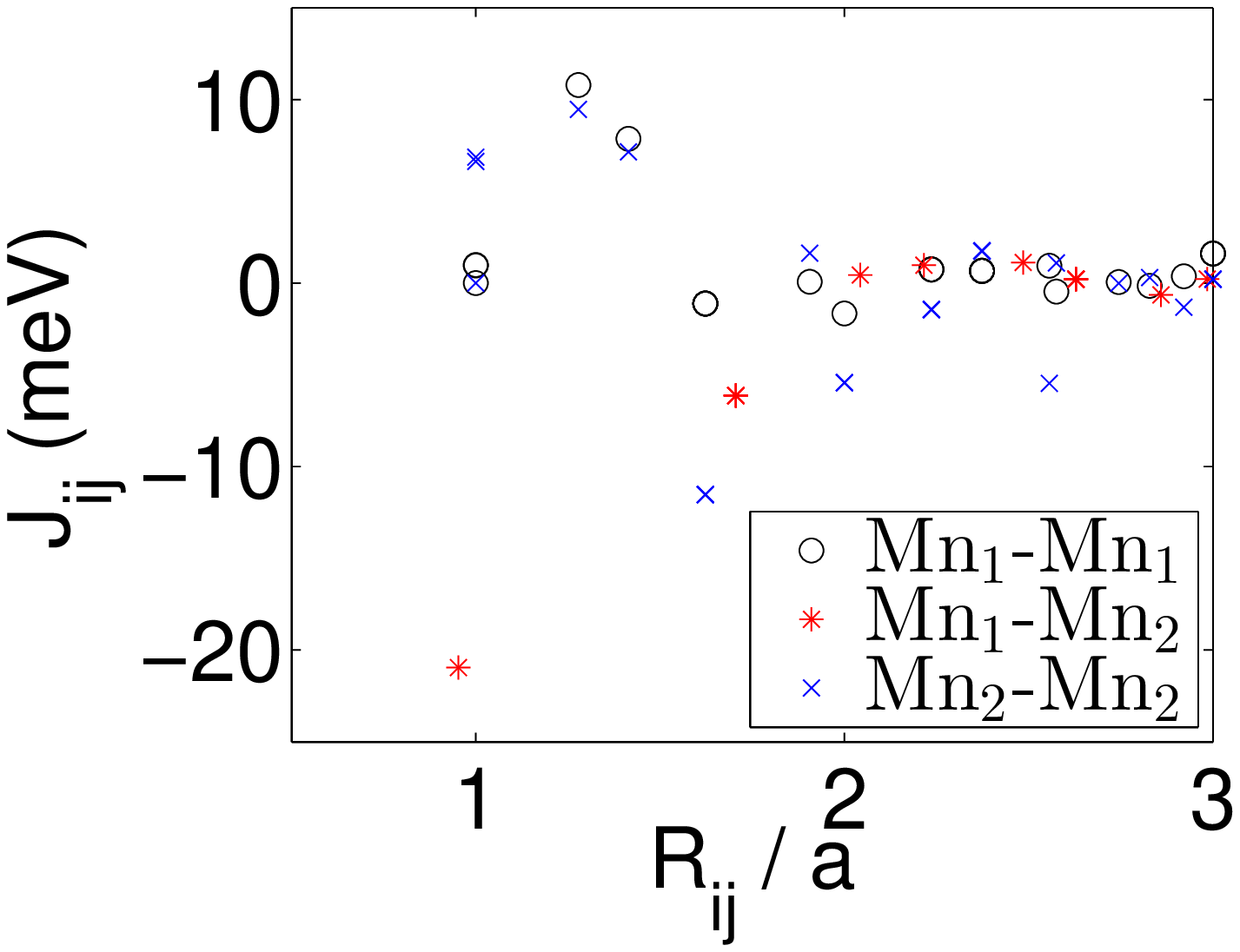}} \\
	\subfloat[MnGa]{\label{fig:MnGaJ}\includegraphics[width=0.25\textwidth]{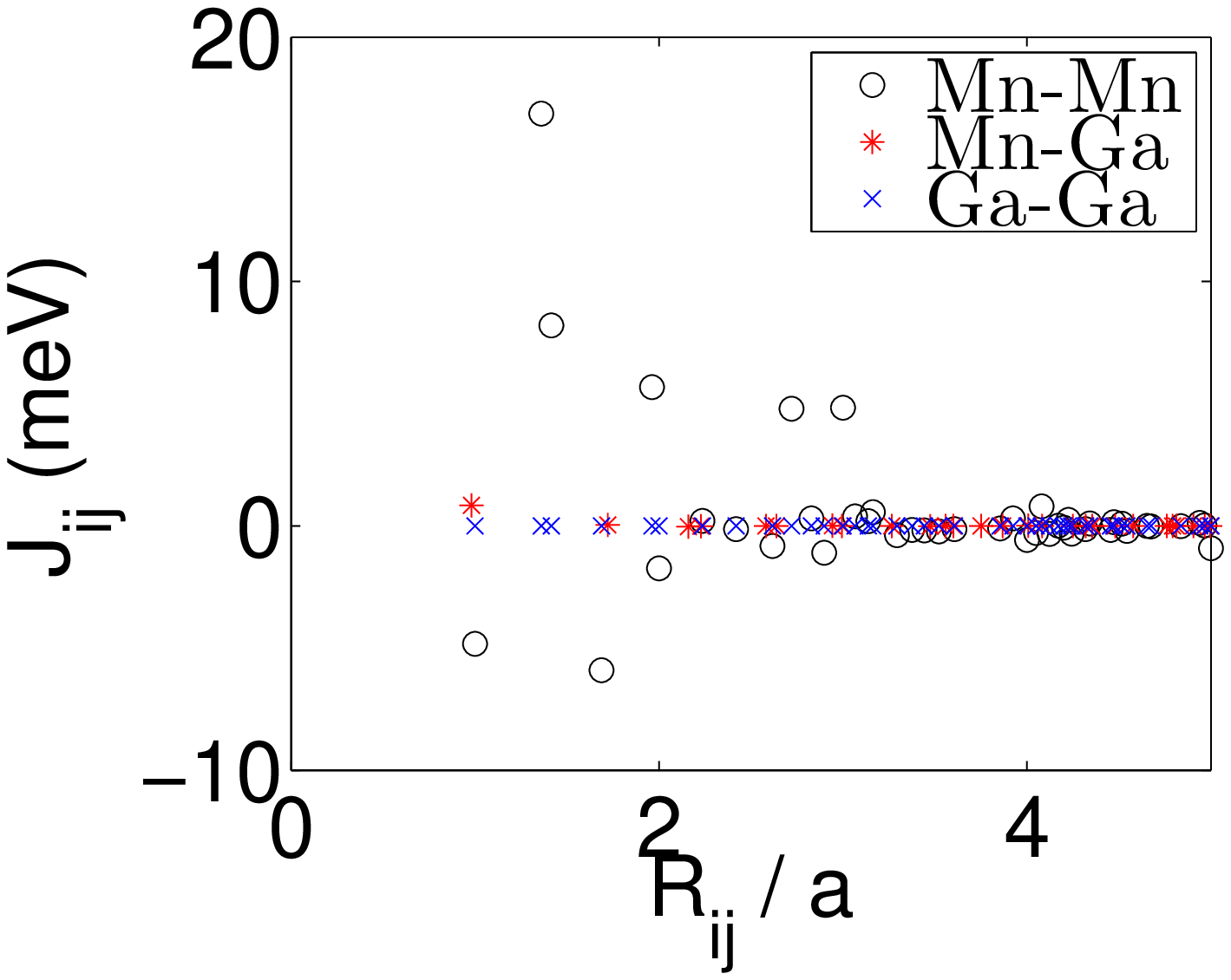}}
	\subfloat[Mn$_{1.20}$Ga$_{0.80}$]{\label{fig:Mn120Ga80J}\includegraphics[width=0.25\textwidth]{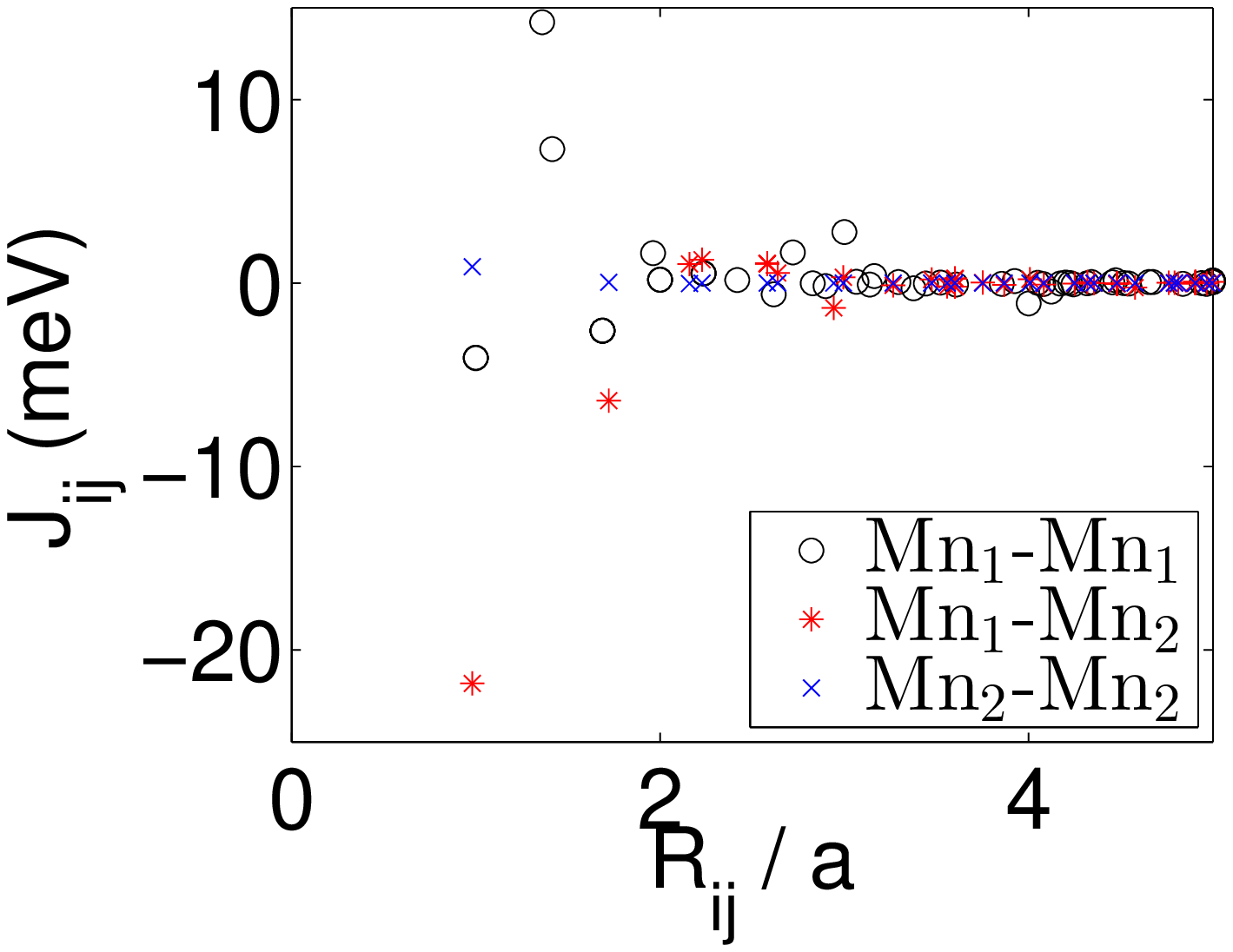}}
	\caption{Atomic distance dependence of exchange parameters $J_{ij}$.}
	\label{fig:Jij_MnAl}
\end{figure}

For stoichiometric and perfectly ordered MnAl, the Monte Carlo simulations show that an antiferromagnetic ordering is prefered over a ferromagnetic order. Competing antiferromagnetic exchange interactions can sometimes infer complex non-collinear ground states, but for MnAl, no such tendency was found from the Monte Carlo simulations. The preference of antiferromagnetism in MnAl can be qualitatively understood if one looks at the exchange interactions as a function of the distance between atoms. Fig.~\ref{fig:MnAlJ} shows that the Mn-Mn interactions have quite strong antiferromagnetic interactions. When introducing Mn also in the second sublattice, one can observe reduction of the antiferromagnetic coupling between Mn atoms in the first sublattice while there is a strong antiferromagnetic coupling between Mn atoms in different sublattices, as seen in Fig.~\ref{fig:Mn114Al86J}. This stabilizes a ferrimagnetic state with Mn atoms in different sublattices having moments in opposite directions, giving a total magnetic moment reduced to $1.98\mu_\text{B}/\text{f.u.}$, but a considerable critical temperature of $T_\text{C} = 670~\text{K}$ in Mn$_{1.14}$Al$_{0.86}$. Experimentally it has also been reported that increased Mn content can cause increased $T_\text{C}$ to, for example, $T_\text{C} = 655~\text{K}$ for Mn$_{1.08}$Al$_{0.92}$\cite{Zeng2007}. 

In MnGa only a weak ferromagnetism with very low $T_\text{C}$ around $80~\text{K}$ was found. Similar behaviour as for MnAl is observed in the $J_{ij}$ of MnGa, as shown in Fig.~\ref{fig:MnGaJ}-\ref{fig:Mn120Ga80J}. Again, increased Mn content yields a higher $T_\text{C}$ and antiferromagnetic coupling between the Mn sublattices yields a reduced total moment. We find, for Mn$_{1.20}$Ga$_{0.80}$,  $T_\text{C} = 690~\text{K}$ and the saturation magnetization reduced by almost $30\%$ to $M_\text{S} = 0.66~\text{MA/m}$. Experimentally it has been reported that pure 1:1 stoichiometric MnGa is not stable, while with 55-60 at.\% Mn it is, and in this range $T_\text{C}$ increases and $M_\text{S}$ decreases with increasing Mn content\cite{Tanaka1993}, which is consistent with our calculations of substitutional disorder. Mn$_{1.18}$Ga$_{0.82}$ has experimentally been reported to show $T_\text{C} = 646~\text{K}$ and $M_\text{S} = 0.39~\text{MA/m}$ at room temperature\cite{Tanaka1993}. At lower Mn content, with around 10-12\% excess Mn, we find more complicated magnetic structures from MC at low temperatures which yields a total moment lowered by about a factor half. Such drastic decreases of moment have also been reported experimentally, although for a bit higher Mn content\cite{Lu2006}. The MAE of Mn$_{1.20}$Ga$_{0.80}$ is, according to SPR-KKR calculations, as large as $429~\mu\text{eV}/\text{f.u.}$

\begin{figure}[hbt!]
	\centering 
	\subfloat[FeNi]{\label{fig:FeNiDOS}\includegraphics[width=0.45\textwidth]{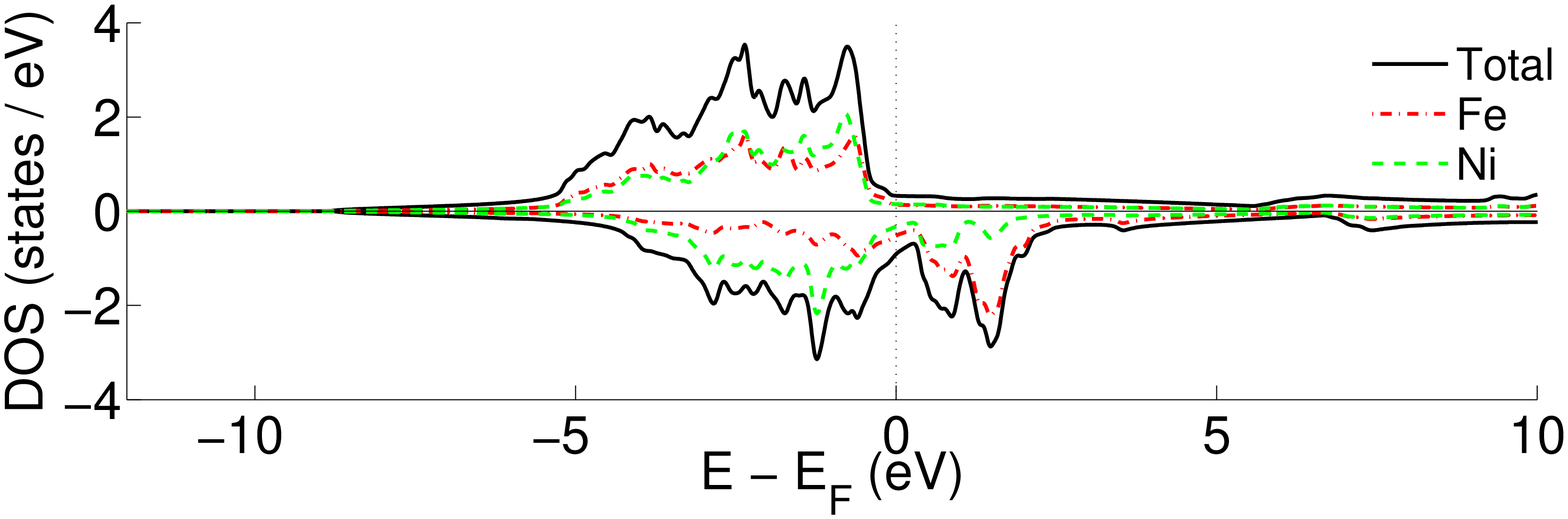}} 	\\
	\subfloat[CoNi]{\label{fig:CoNiDOS}\includegraphics[width=0.45\textwidth]{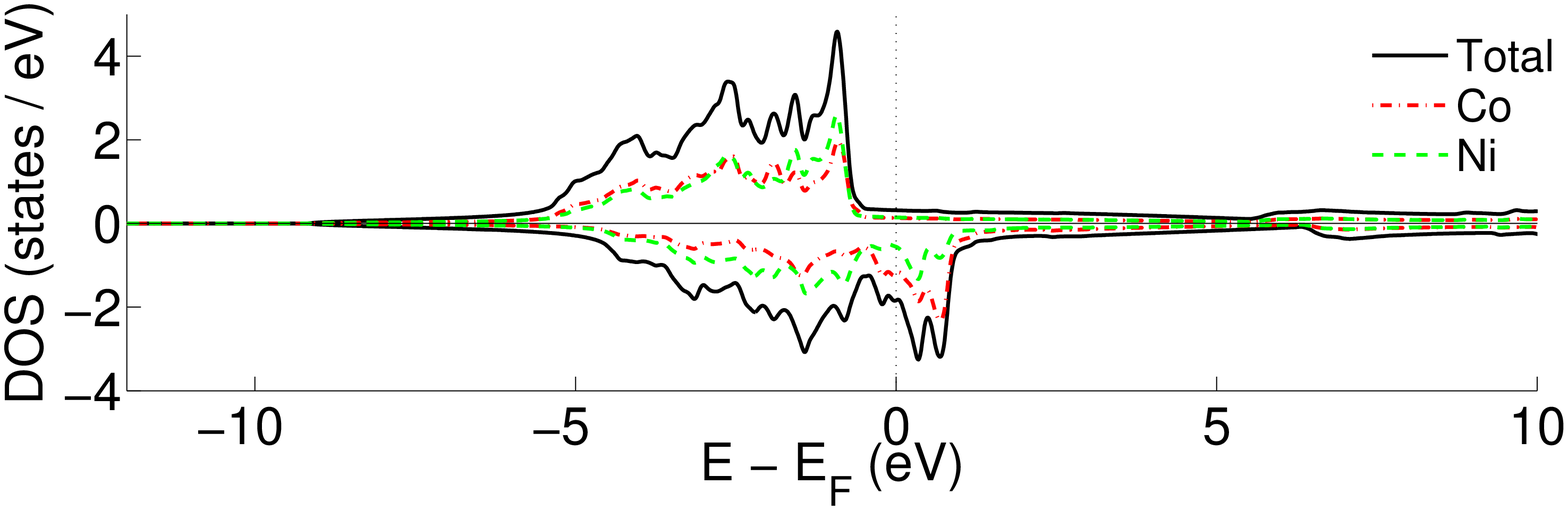}} 	\\
	\subfloat[MnAl]{\label{fig:MnAlDOS}\includegraphics[width=0.45\textwidth]{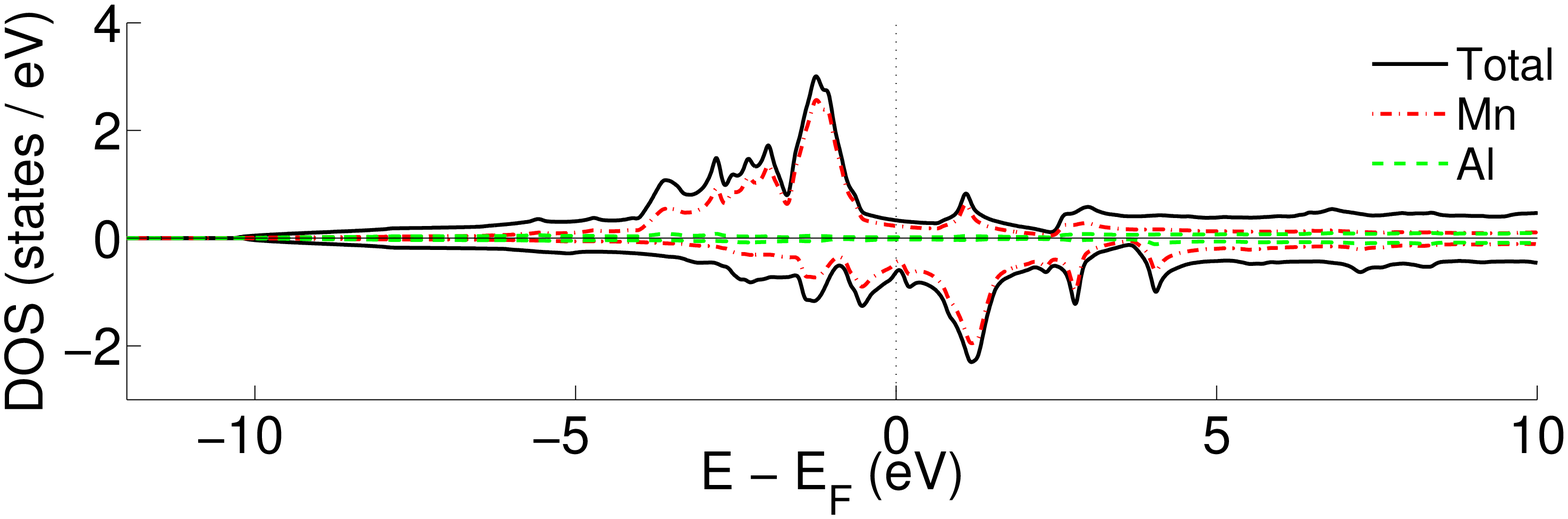}} 	\\
	\subfloat[MnGa]{\label{fig:MnGaDOS}\includegraphics[width=0.45\textwidth]{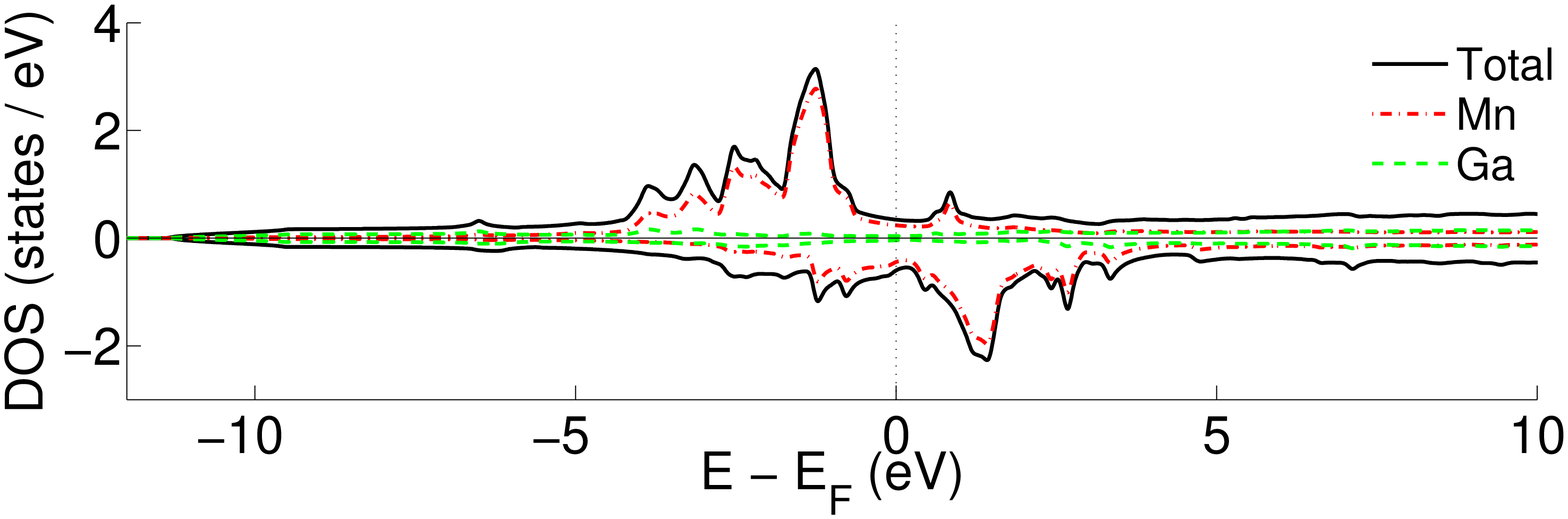}} 	\\
	\caption{Spin polarized density of states.}
	\label{fig:DOS}
\end{figure}

Fig.~\ref{fig:DOS} shows spin-polarized density of states (DOS) around the Fermi energy, calculated in WIEN2k, for the studied stoichiometric compounds. All the plots display a behaviour with clear exchange splitting as expected for ferromagnetic metals and are also in accordance with preceding results for those cases which have been previously studied\cite{Miura2013,Park2010,Sakuma1998}, i.e. FeNi, MnAl and MnGa. The DOS for Ni is seen to be very similar in FeNi and CoNi, although, a small peak just below $-1~\text{eV}$ in the spin down DOS of Ni in FeNi, not present in CoNi, explains a slightly reduced moment of the Ni atom in FeNi compared to that in CoNi. The DOS of MnAl and MnGa are very similar with a pronounced ferromagnetic exchange splitting of just over $2~\text{eV}$ on the Mn atom while Ga and Al exhibit very flat DOS around $E_\text{F}$. One then expects overall similar magnetic properties of the two compounds but at the same time MnGa shows a considerably larger MAE, which is likely due to stronger spin-orbit interaction induced by the Ga atom relative to Al\cite{Andersson2007}. Another possible reason for increased MAE in MnGa, relative to MnAl, is increased $\frac{c}{a}$ which might allow for better localization of d-orbitals along the z-axis, but this is not likely the cause as there is not a significant difference in the occupation of d-orbitals in the two compounds. No significant qualitative changes occur in the DOS when introducing disorder or off stoichiometric compositions. 

In conclusion, the magnetic properties of L1$_0$ binary alloys FeNi, CoNi, MnAl and MnGa have been investigated, systematically and comprehensively, using two different DFT methods. Furthermore, the Curie temperatures have been studied in order to have a complete picture of the three properties $M_\text{s}$, MAE and $T_\text{C}$ which are important in permanent magnet applications. Three of the studied compounds, namely CoNi, MnAl and MnGa, exhibit MAE in the order of $1~\text{MJ/m}^3$ or higher, which is impressive for rare-earth and platinum free materials. Furthermore, all the compounds show Curie temperatures in the order of $600~\text{K}$ or higher, allowing them to be used in permanent magnet applications above room temperature, although we have shown that for Mn-based compounds it is of importance to increase the Mn-content in order to obtain high Curie temperatures. We have also explained the experimentally observed effect of reduced moment in Mn rich compounds due to antiferromagnetic coupling between Mn atoms in the two sublattices. In addition, we have shown that, for FeNi and CoNi, it is of great importance to obtain a high degree of chemical ordering as both MAE and $T_\text{C}$ are reduced by substitutional disorder. 

We thank Olle Eriksson for discussions and feedback. We acknowledge support from EU-project REFREEPERMAG, eSSENCE and Swedish Research Council (VR). Swedish National Infrastructure for Computing (SNIC) and NSC Matter are acknowledged for providing computational resources.

\bibliography{literature}{}
\bibliographystyle{apsrev}

\end{document}